# DETRENDED FLUCTUATION ANALYSIS OF AUTOREGRESSIVE PROCESSES

Vasile V. Morariu[1], Luiza Buimaga-Iarinca[1], Călin Vamoş[2], Ştefan M. Şoltuz[2]

[1]Department of Molecular and Biomolecular Physics, National Institute of R&D for Isotopic and Molecular Technology, R-400293  Cluj-Napoca, P.O.Box 700  Romania, E-mail: vvm@L40.itim-cj.ro,  [2]"T. Popoviciu" Institute of Numerical Analysis, Romanian Academy, P.O. Box 68, 400110 Cluj-Napoca, Romania

Autoregressive processes (AR) have typical short-range memory. Detrended Fluctuation Analysis (DFA) was basically designed to reveal long range correlation in non stationary processes. However DFA can also be regarded as a suitable method to investigate both long-range and short range correlation in non-stationary and stationary systems. Applying DFA to AR processes can help understanding the non uniform correlation structure of such processes. We systematically investigated a first order autoregressive model AR(1) by DFA and established the relationship between the interaction constant of AR(1) and the DFA correlation exponent. The higher the interaction constant the higher is the short range correlation exponent. They are exponentially related. The investigation was extended to AR(2) processes. The presence of a distant positive interaction in addition to a near by interaction will increase the correlation exponent and the range of correlation while the effect of a distant negative interaction will decrease significantly only the range of interaction. This analysis demonstrate the possibility to identify and AR(1) model in an unknown DFA plot or to distinguish among AR(1) and AR(2) models. The analysis was performed on medium long series of 1000 terms.

## 1. Introduction

Many natural processes can be described by stochastic models. The time or scale dependent correlation characteristics of such processes is characterized by the autocorrelation function *C(n)* of two points in time with *lag n.* They may involve either long range correlation or short range correlation characteristics. The decaying of *C(n)* for a long range correlated process lacks a characteristic time. In fact the de-correlation time is infinite. A well known class of long range correlation processes is *1/f* phenomena (*f* is frequency) [1]. On the other hand the decay of short range correlation processes is described by an exponential function and therefore they exhibit a finite time scale. Typical examples of short range memory are the autoregressive processes (AR) [2-3].

Detrended fluctuation analysis (DFA) was a method basically designed to investigate long range correlation in non stationary series [4-6]. DFA produces a fluctuation function *F(n)* as a function of lag *n.* The plot of log*F(n)* vs log *n* is a straight line if long range-correlation is obeyed. The slope of the fluctuation plot is the so called α scaling exponent which has values 0.5 and 1.5 for random (uncorrelated) series and Brownian noise respectively. However, very often, in practice the DFA plot is not a straight line. The plot, in fact, does not define a single long range correlation exponent but instead two or more correlation ranges. Each exponent usually covers at most one order of magnitude scale. The simplest interpretation of such plot is that it represents domains of short range correlation. Therefore DFA can be used to investigate the short range correlation behavior of a process. An appropriate systematic way to investigate the behavior of DFA in case of short range memory is to analyze AR processes. It is of direct interest to investigate the characteristics of the DFA plot or in other words the short range correlation characteristics. The purpose of this work is to perform DFA on the first order and second order autoregressive models known as AR(1) and AR(2) respectively. The main interest is to look for the relationship between the correlation exponent and the characteristic parameters of the AR(1) and AR(2) models. This may help diagnose a fluctuation function in terms of short range memory characteristics.

**2. Autoregressive Models**

An AR(1) model is given by the equation:

$$X_t = \varphi X_{t-1} + \varepsilon_t \qquad (1)$$

where $\varepsilon_t$ is a white noise process with zero mean and variance $\sigma^2$, while $\varphi$ is a parameter. The parameter vales $\varphi$ have to be restricted for the process to be stationary which means that $|\varphi|<1$. If $\varphi=1$ then $X_t$ can also be considered as a random walk.

AR(1) model may apply to temporal or spatial processes and the significance of $\varphi$ can be better understood when applied to such particular processes. For example in the case of a temporal process the parameter $\varphi$ can be understood in terms of a relaxation time $\tau$ determined by [7]:

$$\tau = -\frac{1}{\log(\varphi)} \qquad (2)$$

Parameter $\varphi$ can alternatively be regarded as the strength of interaction among the terms $X_i$ [7]. Obviously the more distant the terms of the series the lower is the correlation. Regardless of a temporal or spatial process the parameter $\varphi$ can be understood as the scale of short range memory of the system.

Higher order models AR(p) are characterized by $i$ parameters of $\varphi_i$ which indicate the strength of interaction between the first and the second term, between the first and the third term and so on. The model is given by the equation:

$$X_t = \sum_{i=1}^{p} \varphi_i X_{t-i} + \varepsilon_t \qquad (3)$$

with $\varphi_i$ parameters where $i = 1,...,p$.

AR models have successfully been applied to astrophysical and psychological data [8-9]. More recently we found that various biophysical phenomena can be well described by AR(1) models or by higher order AR(p) models [unpublished work]. They include the structure of proteins, flickering of the red blood cells and random number generation by human subjects. It was however felt that a systematic DFA of a short range memory model is needed to better understand how the correlation and the scale of memory are related.

**3. Detrended fluctuation analysis of AR(1) model**

Autoregressive series of 1000 terms were generated with a program written in MATLAB. The DFA method involves an integration of the series which is further divided into boxes of equal size *n*. In each box the integrated series is fitted by using a polynomial function The integrated series is detrended by subtracting the local trend in each box. For a given box size *n* a detrended fluctuation which is called the local trend.

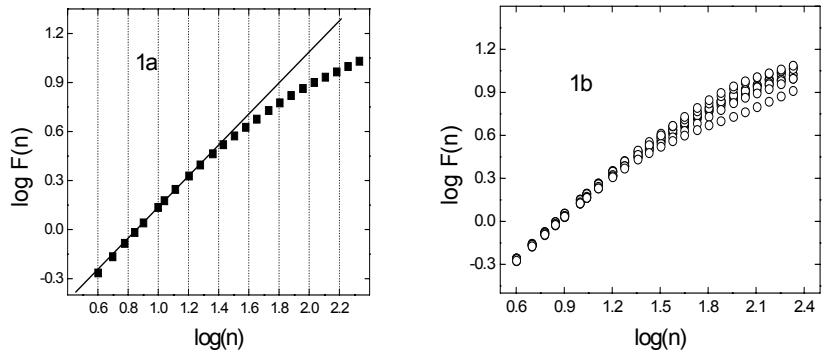

Fig.1. a) Detrended fluctuation analysis of AR(1) model for φ=0.7. A correlation exponent α=0.97 is calculated from the slope of straight line approximation of the first part of the plot. This correlation exponent is characteristic for the short range interval 0.6<log (n)<1.4. b) The same plot generated with ten different random series.

function is calculated then the root mean square fluctuation *F(n)* is obtained. Finally a DFA plot log *F(n) vs.* log (*n*) is obtained. The slope of the plot represents the correlation exponent *α*. In our analysis we used the standard DFA-1 method which means that the local trend was fitted with a first degree polynomial. An example of DFA plot is illustrated in figure 1a. It can be seen that within range 0.6 < log (n) <1.4 the system is characterized by a correlation exponent $α_1$ which describes a short range memory. This result was obtained from a single generated series of data. If different AR(1) series are generated by starting from different random series then a bunch of curves which diverge at higher *n* values resulted (fig.1b).

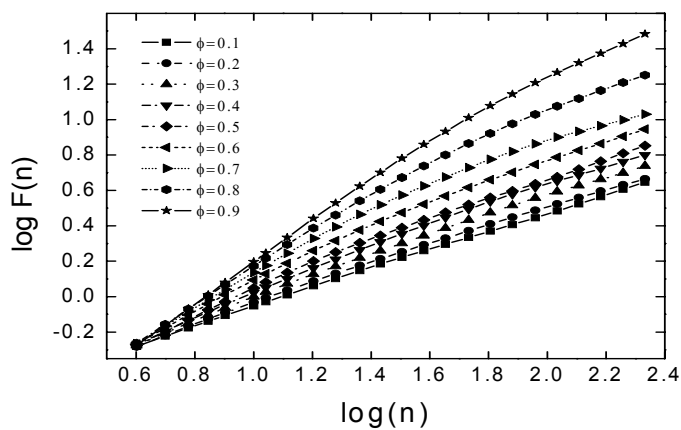

Fig.2. Averaged detrended fluctuation analysis plots of AR(1) models for different values of parameter φ. The plot illustrate the short range correlation of the AR(1) model as the slope of the plot (correlation exponent) gradually decreases at higher *n*.

Consequently each case of AR(1) model for a given φ was averaged out for ten generated series. Averaged plots are represented in fig.2 for different values of φ. The correlation is limited to ranges shorter than about one order of magnitude of *n*. The correlation decreases gradually at distances longer than about one order of magnitude. Stronger correlation of the DFA plot is visible as a higher slope (higher $\alpha_1$ value) for higher values of φ. The upper part of the DFA plots can also be described by a correlation exponent $\alpha_2$ which is smaller than $\alpha_1$. The standard deviation for the end of each curve varied between 0.04 and 0.1.

Further we illustrate how the correlation exponent $\alpha_1$ depends on the value of φ (fig. 3a). We notice that the short range correlation exponent $\alpha_1$ is closely correlated to the AR(1) parameter φ. It should be stressed that $\alpha_1$ exponent is a correlation property while φ is a scale parameter describing either a characteristic time or the strength of interaction among the terms of the series. The range of short range correlation Δlog *(n)*, *i.e.* the linear domain on which exponent $\alpha_1$ is defined, as function of φ, is illustrated in fig. 3b. The higher the value of φ the longer is the range of the short range correlation.

The upper part of the DFA plot (fig. 1) can also be characterized by a slope which corresponds to a $\alpha_2$ exponent. Its dependence on the value of parameter φ is illus- illustrated in figure 4. It shows that the upper part of the DFA plot of an AR(1) model is practically uncorrelated ($\alpha_2$=0.5). Correlation still persists at φ≥0.8 where $\alpha_2 \geq 0.6$. On the other hand this range of correlation decreases with increase of φ (fig.4b) as opposed to behavior of the range of correlation for $\alpha_1$ (fig. 3b).

These examples show that an AR(1) process can be conveniently characterized by, at least, a pair of $\alpha_1$ and Δlog*(n)* values. Consequently these data can be easily used to identify an AR(1) process or exclude it when confronted to a DFA plot of unknown origin. It would probably be safer to consider $\alpha_1$ and $\alpha_2$ values and Δlog(n)$_1$ and Δlog(n)$_2$ respectively for a DFA plot of an unknown origin in order to identify or reject the validity of AR(1) model. If all four values point to the same φ value then the data conform to an AR(1) model.

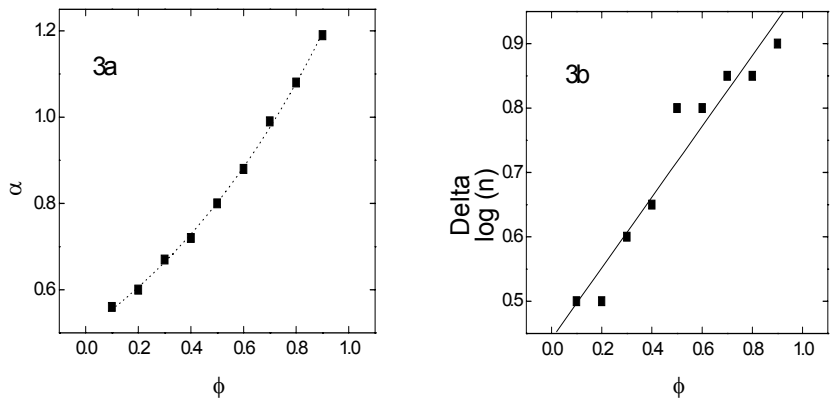

Fig.3. a) The influence of AR(1) parameter φ on the short range correlation DFA exponent $\alpha_1$. The data are fitted with an exponential. b). The range of short range correlation *Delta(log n)* in the DFA plot corresponding to exponent $\alpha_1$ for different values of the AR(1) parameter φ. The data are fitted by a straight line.

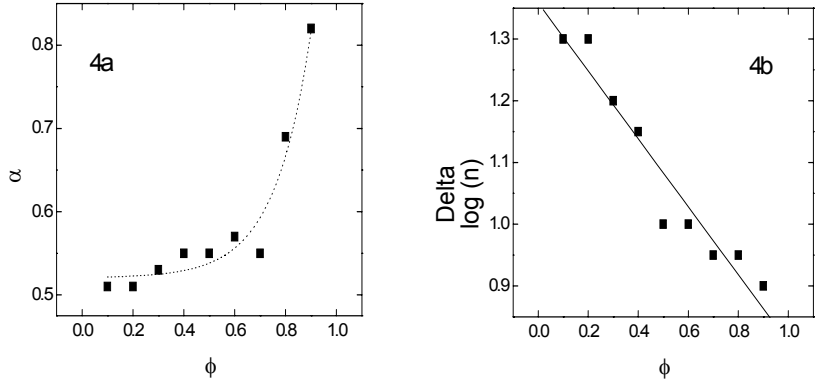

Fig. 4. a) The influence of parameter φ on the correlation exponent $α_2$ of the upper part of the DFA plot for an AR(1) model. The dependence is exponential. b) The range of short range correlation for $α_2$ exponent.

## 4. Second order AR model

It is not the purpose of this article to discuss a systematic behavior of DFA for higher order AR models. This is a matter of a very large number of possibilities. However we provide examples of AR(2) models which may suggest how to distinguish among AR(1) and higher order AR(p) models. DFA of two AR(2) models compared to a related AR(1) model are shown in fig.5. In our example the AR(2) model is generated with the same $φ_1=0.6$ (as for AR(1)) and different $φ_2$ values ($φ_2=0.3$ and $φ_2=-0.3$ respectively).

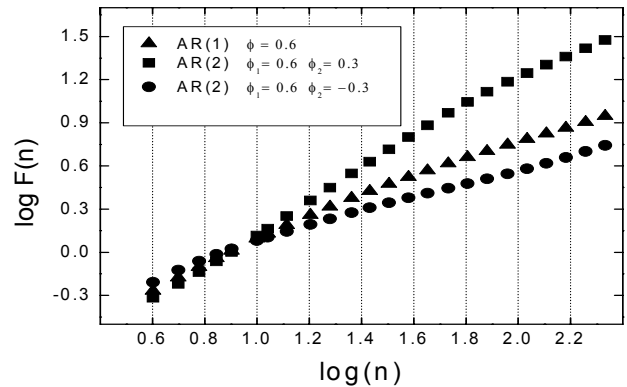

Fig.5. Example of DFA plots for AR(2) model compared to AR(1).

The results differ in respect to both correlation exponent α and the range of correlation $\Delta \log(n)$. The presence of the additional interaction $φ_2=0.3$ significantly increases the range of correlation from $\Delta\log(n)=0.8$ to 1.25 and the correlation exponent α from 0.85 to 1.15. On the other hand $φ_2=-0.3$ decreases only slightly the correlation from α=0.85 (for AR(1)) to α=0.8 but the effect is stronger on the range of correlation which decreases from Δlog(n)=0.8 to 0.3.

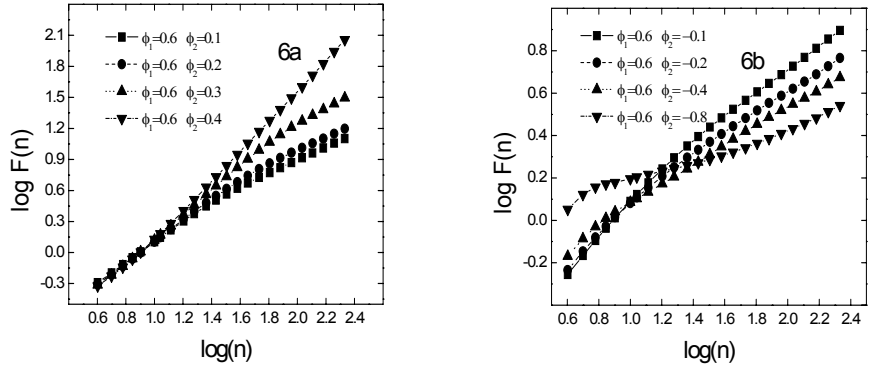

Fig.6. a) The influence of a positive value of $\varphi_2$ on the DFA plot of an AR(2) model with $\varphi_1$=0.6. b) The influence of a negative value of $\varphi_2$ on the DFA plot of an AR(2) model with $\varphi_1$=0.6.

The influence of $\varphi_2$ on the DFA plots for an AR(2) model with $\varphi_1$=0.6 is further illustrated in fig.6. The effect is obviously different for positive values of $\varphi_2$ and negative values respectively. An analysis in terms of short range correlation exponents would be more complicated than in the case of AR(1). The effect of a positive $\varphi_2$ does not change qualitatively the DFA plot against an AR(1) model. However values of $\varphi_2$ higher than about 0.4 increased the correlation to the limit of a random walk. A careful inspection of these DFA plots may qualitatively identify an AR(2) process or distinguish an AR(1) against an AR(2) process. Suppose the DFA plot of an unknown origin provides a value $\alpha_1$=0.8. According to the DFA plot in fig.3a this corresponds to $\varphi$=0.5 for an AR(1) model. The corresponding range of correlation for AR(1) is $\Delta\log(n)$=0.72 (fig.3b). If the unknown DFA plot define a similar range of correlation it is likely that the short range correlation property of the series can be described by an AR(1) model. Obviously an AR(2) model will not fulfill this condition as the range of correlation is different.

      Other characteristics of AR(2) such as the effect of a negative value of $\varphi_2$ can be easily recognized by a significant inflexion of the DFA plot or by a less important inflexion at positive values of $\varphi_2$. Further it shows that the local correlation varies continuously in a more complicated way for AR(2) with negative $\varphi_2$ values of about 0.7-0.8. Analysis in terms of short range correlation and range of correlation gradually changes to local correlation properties as $\varphi_2$ becomes more negative (fig.6b).

## 5. Conclusions
DFA correlation exponent is nonlinearly related to the AR(1) interaction parameter $\varphi$ and the range of correlation is linearly related to the same parameter. DFA of first order autoregressive processes shows that the correlation exponent and the range of correlation represent a pair of characteristic data for a given AR(1) process. Consequently an unknown AR process can be identified as an AR(1) process by confronting its characteristics with theoretical values. A second order AR(2) model revealed that a positive distant interaction $\varphi_2$ increased both the correlation exponent and the range of correlation compared to and AR(1) with the same nearby interaction constant $\varphi_1$. A negative distant interaction $-\varphi_2$ significantly decreased the range of correlation. A careful qualitative analysis of the DFA plots may identify an AR(2) process or distinguish among AR(1) and AR(2) processes respectively.

**Acknowledgements**

Funding of the work was provided by the Romanian Authority for Scientific Research.

(2)